\documentclass[preprint,12pt]{elsarticle}

\usepackage{amssymb}
\usepackage{amsmath}
\usepackage{float}
\usepackage{bm}
\usepackage{graphicx}
\usepackage{booktabs}
\usepackage{multirow}
\usepackage{tabularx}
\usepackage{subcaption}
\usepackage{dsfont}
\usepackage{soul, color}
\usepackage{makecell}
\usepackage{placeins}
\usepackage{dsfont}
\usepackage[nocomma]{optidef}
\usepackage{siunitx}
\usepackage{xurl}

\newcommand{\aFRR}{\text{aFRR}}
\newcommand{\mFRR}{\text{mFRR}}
\newcommand{\B}{\mathrm{B}}
\newcommand{\onemin}{\SI{1}{\minute}}
\newcommand{\fifteenmin}{\SI{15}{\minute}}
\newcommand{\onehour}{\SI{1}{\hour}}

\journal{Renewable Energy}

\begin{document}

\begin{frontmatter}

\title{Every minute counts: Wind farm power control during imbalance settlement using distributional soft actor-critic}

\author[1]{Max Bruninx\corref{cor1}}
\ead{max.bruninx@vub.be}
\author[2]{Seyed Soroush Karimi Madahi}
\author[1]{Timothy Verstraeten}
\author[1]{Jan Decuyper}
\author[2]{Chris Develder}
\author[1]{Ann Nowé}
\author[1]{Jan Helsen}

\cortext[cor1]{Corresponding author}

\affiliation[1]{organization={Vrije Universiteit Brussel},
            city={Brussels},
            country={Belgium}}
\affiliation[2]{organization={Ghent University},
city={Ghent},
country={Belgium}}

\begin{abstract}
Wind farms can provide flexibility by curtailing their output during periods of excess generation in the power grid. One way to achieve this is through implicit balancing, where curtailment is performed in anticipation of a negative imbalance price. By doing so, the wind farm helps reduce system imbalance while also lowering its own costs. However, implicit balancing is challenging, since imbalance prices are hard to predict and can also be affected by the control actions taken by the farm. To address both challenges, we use distributional soft actor-critic~(distSAC) to learn a control policy that adjusts the power setpoint every minute. Using real data of the Belgian grid and an offshore wind farm, we show that our method can turn the imbalance costs caused by forecast errors into a small revenue. Moreover, distSAC outperforms a price threshold controller, often used in industry, and performs slightly better than soft actor-critic (SAC). Unlike SAC, distSAC can also learn risk-averse policies by optimizing a risk measure, such as conditional value at risk. These risk-averse policies were found to better account for their price impact, reducing overshoots by 16-27\% while achieving similar cost savings. Last, we find that 1-minute setpoints lead to lower imbalance costs than 15-minute setpoints, regardless of the control policy.
\end{abstract}


\begin{keyword}
Wind farm power control \sep renewable flexibility \sep imbalance settlement \sep distributional reinforcement learning
\end{keyword}

\end{frontmatter}

\section{Introduction}

The growth in global renewable energy capacity \cite{IEA_GlobalEnergyReview2025} makes it more challenging for the Transmission System Operator (TSO) to maintain system balance. Consequently, the need for flexibility, i.e., assets that can resolve deviations between electricity supply and demand, is increasing. In addition to conventional flexible assets, such as battery storage, controllable renewable assets could also offer such flexibility services \cite{elia2026flexibility, liu2023evolution, todd2026flexible}. Specifically, these assets can reduce their power output when the system has an energy surplus, a process known as \textit{curtailment}. By doing so, they can improve their overall revenue while also helping to address one of the main challenges of integrating renewable energy into the power grid.

Flexibility services can be provided to the grid either explicitly, through ancillary services, or implicitly, during imbalance settlement. The former requires offering the available flexible capacity to the reserve markets beforehand and following the setpoints sent by the TSO in real-time. The latter, in contrast, implies controlling the power output of the asset based on the imbalance price. Since single imbalance pricing has become prevalent in Europe \cite{acer2020harmon}, Balance Responsible Parties (BRPs) can be rewarded for deviations from their trading schedule in a way that contributes to system balance, also called \textit{implicit balancing}. However, it is important to note that not all European countries abide by the same regulation related to this behavior. For example, in Belgium the TSO explicitly encourages implicit balancing \cite{elia2025flexibility}, whereas in Germany it is prohibited \cite{eicke2021electricity}. The reason for these distinct views is mainly related to the trade-off between sacrificing explicit control and the risk of overshoots versus the potential cost savings and reduced entry barriers for flexible assets \cite{bruninx2026herd}.

In this work, we consider the control of an offshore wind farm that participates in implicit balancing through curtailment. A gap in the literature exists in this regard, as prior research has mainly focused on implicit balancing strategies for storage systems. The curtailment of renewables, however, constitutes a different control problem, since the available capacity depends on the weather and cannot be shifted to a later point in time. The choice for implicit balancing is motivated by the low entry barrier compared to explicit balancing, as any BRP can participate by controlling their assets in real-time. Additionally, implicit balancing offers potential cost reductions and can, therefore, be seen as an interesting new revenue stream. This is particularly important for offshore wind farms given their declining subsidies~\cite{totalenergies2024negative,rvo2024negative}. Nonetheless, it remains a challenging problem since the actual imbalance price is only known after settlement and prices are difficult to forecast even in the short term \cite{bottieau2019very}. Furthermore, the installed capacity of offshore wind farms is large relative to most system imbalances, such that controlling these assets could significantly affect the imbalance price. 

To deal with these challenges, we resort to model-free reinforcement learning (RL), which allows a control policy to be learned directly from interaction with the environment. The policy thereby does not need to explicitly model the underlying non-linear dynamics of the imbalance price \cite{bruninx2026herd}, but instead captures their effects through experience. Specifically, we adopt distributional soft actor-critic (distSAC), an extension of soft actor-critic (SAC) to distributional reinforcement learning \cite{bellemare2017distributional}. The main idea of distributional RL is to estimate the entire return distribution rather than employing the expected value or Q-value. This was found to stabilize learning \cite{bellemare2017distributional} and enable risk management \cite{theate2023risk}. Additionally, we argue that a distributional perspective is a better fit to our setting given the stochastic nature of the implicit balancing problem. 

The main contributions of our work can be summarized as follows:
\begin{itemize}
    \item To the best of our knowledge, we are the first to show that the curtailment of renewables without co-located storage can provide real-time flexibility through implicit balancing.
    \item We develop a realistic simulation environment that combines our minute-level imbalance price simulator~\cite{bruninx2026herd} with supervisory control and data acquisition (SCADA) data of Norther, an offshore wind farm with a capacity of 370 MW.
    \item We adopt distributional soft actor-critic to learn both risk-neutral and risk-averse control policies for implicit balancing.
\end{itemize}


The results show that distSAC can turn the imbalance costs caused by day-ahead forecast errors into a small revenue, whereas a price threshold controller reduces these costs by only about half. Furthermore, distSAC achieves slightly better performance than SAC, while also supporting risk management. The risk-averse distSAC policies reduce the number of overshoots by 16-27\% and still attain an imbalance cost reduction comparable to that of the risk-neutral policy. Last, refining the setpoint resolution from 15 minutes to 1 minute lowers imbalance costs for every evaluated control policy. This finding is of practical importance, as 15-minute setpoints are commonly adopted in both industry and the academic literature~\cite{bottieau2019very, lago2020optimal, smets2023strategic, smets2025value, wessel2024risk, madahi2027neural, rasic2025safe}.

The remainder of the paper is structured as follows: Section~\ref{sec:related_work} discusses related work and Section~\ref{sec:problem_setting} describes the problem setting. Sections~\ref{sec:sac} and~\ref{sec:distsac} introduce soft actor-critic and distributional soft actor-critic, respectively. Section~\ref{sec:results} presents numerical results to validate our method. Last, Section~\ref{sec:discussion} concludes the paper with a discussion of the findings and directions for future work.

\section{Related work}
\label{sec:related_work}

There are two categories of research related to increasing revenue through imbalance settlement. The first category, which we will call \textit{trading arbitrage}, aims to construct short-term trading strategies that benefit from imbalance settlement by buying or selling electricity when the day-ahead or intraday price is, respectively, lower or higher than the expected imbalance price. The second category focuses on \textit{real-time asset control}, where the goal is to adjust power production and consumption based on the system needs. Our study falls into the second category, but the two categories are closely related and strategies from both can be combined \cite{wessel2024risk, ally2025day}.

\subsection{Trading arbitrage} 
Regarding trading arbitrage, prior work mainly focuses on price-taker arbitrage between day-ahead and imbalance prices. To this end, several studies consider trading strategies for batteries \cite{miljan2022operational} or renewable assets \cite{browell2018risk, heiser2025betting, heiser2026and}. However, the price-taker assumption can negatively impact performance, as shown for wind farms with large installed capacities by \cite{bruninx2025day}. The authors also propose a way to reduce this impact by constraining the bidding strategy based on probabilistic forecasts of wind power generation. While this method was found to reduce the effect of price impact, it does not explicitly account for it. Alternatively, \cite{wessel2024risk} and \cite{zugno2013pool} have proposed to incorporate the price impact in the optimization problem using the merit order of the available reserves. To this end, they use a linear model of the imbalance price formation and incorporate this in the trading arbitrage problem using Karush–Kuhn–Tucker (KKT) conditions. In reality, however, the imbalance price formation is not necessarily linear and this approach may result in a poor approximation of the actual price \cite{madahi2025model}.


\subsection{Real-time asset control}

\begin{table}[b!]
\centering
\resizebox{\textwidth}{!}{%
\begin{tabular}{c ccc cc c c}
\toprule
\multirow{2}{*}{Paper} & \multicolumn{3}{c}{Flexible asset(s)} & \multicolumn{2}{c}{Method} &  \multirow{2}{*}{Price impact} &\multirow{2}{*}{Trading schedule} \\
\cmidrule(lr){2-4} \cmidrule(lr){5-6}& Type & Total capacity & Setpoints & Model-based & Model-free &  &\\
\midrule
\cite{jahromi2026adaptive} & EV & 11 kW & \onemin & & \checkmark &  - &\\
\cite{sales2024efficient} & EV & 330 kW & \onehour & & \checkmark &  - &\\
\cite{rahman2022efficient} & EV & 200 kW & \onehour & & \checkmark &  - &\checkmark \\
\cite{lago2020optimal} & STESS & 300 MW & \fifteenmin & \checkmark & \checkmark &  - &\checkmark \\
\cite{smets2023strategic} & BESS & 2-100 MW & \fifteenmin & \checkmark & &  \fifteenmin &\\
\cite{smets2025value} & BESS & 1 MW & \fifteenmin & \checkmark & &  - &\\
\cite{wessel2024risk} & BESS & 20 MW & \fifteenmin & \checkmark & &  \fifteenmin &\checkmark \\
\cite{pavirani2024frequency} & BESS & 10 MW & \onemin & & \checkmark &  - &\checkmark \\
\cite{hendrickx2026multi} & BESS & 10 MW & \onemin & & \checkmark &  - &\checkmark \\
\cite{madahi2024distributional} & BESS & 1 MW & \onemin & & \checkmark &  - &\\
\cite{pavirani2025system} & BESS & 1 MW & \onemin & & \checkmark &  - &\\
\cite{madahi2025risk} & BESS & 1 MW & \onemin & & \checkmark &  - &\checkmark \\
\cite{madahi2027neural} & BESS & 1-100 MW & \fifteenmin & \checkmark & &  \onemin &\\
\cite{madahi2025model} & BESS & 1-100 MW & \onemin & \checkmark & \checkmark &  \onemin &\\
\cite{bottieau2019very} & BESS & 120 MW & \fifteenmin & \checkmark & &  \fifteenmin &\\
\cite{rasic2025safe} & BESS & 20 MW & \fifteenmin & & \checkmark &  \fifteenmin &\\
\cite{ally2025day} & WF+HESS & 250-450 MW & \onemin & & \checkmark &  \fifteenmin &\checkmark \\
\midrule
Our work & WF & 370 MW & \onemin & & \checkmark &  \onemin &\checkmark \\
\bottomrule
\end{tabular}%
}
\caption{Overview of literature related to real-time asset control during imbalance settlement. The asset types are abbreviated as: electrical vehicle (EV), seasonal thermal energy storage system (STESS), battery energy storage system (BESS), hydrogen energy storage system (HESS) and wind farm (WF).}
\label{tab:literature-review}
\end{table}

Table \ref{tab:literature-review} provides an overview of the literature on real-time asset control during imbalance settlement. The works are categorized based on the flexible asset they consider, including the capacity of the asset and the resolution of the power setpoints. Furthermore, we distinguish between model-based and model-free control methods and assess whether the work includes price impact during imbalance settlement and/or a trading schedule. In terms of assets, all prior literature considers a form of energy storage, with battery energy storage systems representing the largest share. The capacity of these assets differs from kilowatts to hundreds of megawatts and is found to be related to whether or not the method considers the price impact of the asset. Regarding the price impact, the temporal resolution employed to calculate the price must be considered, as it affects how close the estimated price is to reality \cite{bruninx2026herd, madahi2025gaming}. Under a 15-minute resolution, the price is determined at the end of the imbalance settlement period based on the average system imbalance over that period and the reserves required to resolve this imbalance. Under a 1-minute resolution, by contrast, the activation of reserves is based on the minute-level system imbalance and intermediate prices can also be calculated throughout the settlement period. Among the works that model price impact, all except \cite{ally2025day} use the merit order of the reserves. Apart from \cite{sales2024efficient,rahman2022efficient}, the asset controllers attain a setpoint resolution of 15 minutes or less, corresponding to the mandatory time window of the settlement period according to the European Electricity Balancing Guideline \cite{acer2020harmon}. For both the price impact and the setpoint resolution of the controller, inter-quarter hour dynamics are important, as shown by \cite{bruninx2026herd, madahi2025gaming}. A minute-level granularity is therefore preferable, particularly in Belgium, where the TSO publishes new information on the imbalance price and the corresponding system imbalance every minute. Another distinction can be made between model-based and model-free methods. Model-based methods rely on an explicit mathematical model of the system and its environment to compute control actions, whereas model-free methods learn a control policy directly from data or interaction with the environment. Among the model-based methods, optimization over the settlement period is applied in \cite{bottieau2019very}, while model predictive control is employed in all other works. The latter is preferred, mainly because it allows planning over a longer time horizon, which is important for energy storage systems \cite{smets2023strategic}. Among the model-free methods, reinforcement learning is used throughout, with soft actor-critic being the most commonly used algorithm. Last, when a trading schedule is included, a day-ahead position is usually considered, apart from the works of \cite{pavirani2024frequency, hendrickx2026multi}, which use an offer to the Frequency Containment Reserve~(FCR) market.

The closest related work to ours is \cite{ally2025day}, which also considers an offshore wind farm and uses a control methodology based on soft actor-critic. Nonetheless, three distinctions can be made with respect to our study. First, Ally et al.~\cite{ally2025day} approximate the price impact of the control actions with a proxy of the imbalance price based on polynomial regression. As a result, the effect of the merit order within a given settlement period is not accounted for. In reality, however, the merit order has a significant effect on the settlement price, such that the same situation can result in an entirely different price, making the actual imbalance price less predictable compared to their simulator. Second, the authors consider an offshore wind farm co-located with an electrolyzer, which extends the control problem beyond curtailment decisions. The electrolyzer, namely, provides additional flexibility, which can reduce exposure to imbalance costs. As such co-located configurations do not yet exist in Belgium, we restrict our setting to power control of the wind farm, so that our approach is directly applicable to the present situation. Last, our work adopts the distributional variant of soft actor-critic to better capture the uncertainty inherent to imbalance settlement.

\section{Problem setting}
\label{sec:problem_setting}

\subsection{Implicit balancing}

On the day-ahead and intraday markets, energy is sold before generation or consumption takes place. For a wind farm, this implies that its power production is offered to these markets based on forecasts, resulting in a trading schedule~$v_t$. In real time, however, the actual production may deviate from this schedule for three different reasons. First, the intermittency of renewable energy could result in deviations between the power forecasts, which are issued beforehand, and the possible power production~$\overline{P}_t$, which is the maximum power production given the wind conditions. Second, the wind farm operator can apply a farm-level power setpoint~$x_t$ to the turbines in order to reduce, or curtail, the actual power production to~$P_t$. Third, the market participant may offer volumes that deviate from the power forecast, for instance, to engage in trading arbitrage. In this work, we exclude the third reason and, thus, assume that the trading schedule~$v_t$ equals the power forecast. To distinguish between the first two reasons, we define the forecast error as~$\overline{P}_t - v_t$ and the open position as~$P_t - v_t$.

On the other hand, the transmission system operator is responsible for resolving the system imbalance $\mathrm{SI}_t$ in order to stabilize the grid frequency and avoid blackouts. The system imbalance is the difference between the total electricity supply and demand. These imbalances are counteracted by the TSO through the activation of automatic and manual frequency restoration reserves (aFRR/mFRR). The cost of these reserves is then attributed to the generators and off-takers that contributed to the system imbalance\footnote{In practice, the balancing costs are attributed to Balance Responsible Parties which manage a portfolio of generators and off-takers and they attribute these costs to the underlying assets.} through imbalance settlement. Under single imbalance pricing, the TSO pays a reward to (\scalebox{0.8}{$+$}) or receives a penalty from (\scalebox{0.8}{$-$}) the market participants for every imbalance settlement period (ISP). The magnitude of each payment is defined as
\begin{equation}
    \lambda^{\textrm{B}} \sum_{t\in \textit{ISP}} (P_t-v_t)\>\Delta t,
\end{equation}
where $\lambda^{\textrm{B}}$ is the imbalance price of that specific ISP, which is also known as the settlement price.

The objective of implicit balancing for a wind farm is then to adjust the power setpoints $x_t$ such that its imbalance costs are reduced. Simply put, this implies that the farm curtails its power output in case of negative imbalance prices and produces as much as possible when the prices are positive. However, in reality, the problem is more challenging for two reasons. First, the settlement price $\lambda^{\textrm{B}}$ is unknown in advance and difficult to forecast. Therefore, the controller must rely on information published by the TSO during the settlement period, such as an estimate of the imbalance price $\hat{\lambda}^{\mathrm{B}}_t$, to anticipate the settlement price. Second, the open position of the wind farm will also affect the system imbalance and, consequently, the settlement price. Therefore, the controller must carefully select power setpoints that do not cause the system imbalance to change sign, as this event could turn the imbalance payment into a penalty rather than a reward.

\subsection{Imbalance price formation}

In this work, the impact of the control actions on the imbalance price is captured by replicating the imbalance price formation. To this end, we first determine which aFRR and mFRR bids are activated by the TSO to resolve the residual system imbalance $\mathrm{SI}_t$. The activation of these reserves can be formulated as an optimization problem for every optimization cycle $t$:
\begin{mini!}|s|
    {}{
        \sum_{a^+\in \aFRR^+}\mu^{a^+}b^{a^+}_t
        - \sum_{a^-\in \aFRR^-}\mu^{a^-}b^{a^-}_t
    }{\label{opti-obj-given}}{}
    \breakObjective{
        + \sum_{m^+\in \mFRR^+}\mu^{m^+}b^{m^+}_t
        - \sum_{m^-\in \mFRR^-}\mu^{m^-}b^{m^-}_t,
    }\notag
    \addConstraint{\sum_{a^{+}\in \aFRR^{+}}b^{a^{+}}_t}{= V^{\aFRR^{+}}_t: \lambda^{\aFRR^{+}}_t,}{\label{opti-con-given:required_afrr_up}}
    \addConstraint{\sum_{a^{-}\in \aFRR^{-}}b^{a^{-}}_t}{= V^{\aFRR^{-}}_t: \lambda^{\aFRR^{-}}_t,}{\label{opti-con-given:required_afrr_down}}
    \addConstraint{\sum_{m^{+}\in \mFRR^{+}}b^{m^{+}}_t}{= V^{\mFRR^{+}}_t: \lambda^{\mFRR^{+}}_t,}{\label{opti-con-given:required_mfrr_up}}
    \addConstraint{\sum_{m^{-}\in \mFRR^{-}}b^{m^{-}}_t}{= V^{\mFRR^{-}}_t: \lambda^{\mFRR^{-}}_t,}{\label{opti-con-given:required_mfrr_down}}
    \addConstraint{0\leq b^{r}_t\leq B^{r},}{\quad\forall r\in \mathcal{R},}{\label{opti-con-given:individual_capacity}}
\end{mini!}
where $\mathcal{R}$ is the set of all individual balancing bids, each characterized by an activated volume $b^{r}_t$, a price per volume $\mu^{r}_t$ and a total offered capacity~$B^{r}$. Bids are further distinguished by two superscripts: the reserve type, $a$ for aFRR or $m$ for mFRR, and the direction of the balancing energy, denoted \scalebox{0.8}{$+/-$}, indicating, respectively, incremental and decremental energy. In constraints (\ref{opti-con-given:required_afrr_up}) to (\ref{opti-con-given:required_mfrr_down}), the aggregated volumes per reserve type are given by $V^\mathrm{FRR}_t$ and the dual variables $\lambda^\mathrm{FRR}_t$ represent the marginal prices. These aggregated volumes are provided as input to the optimization problem in order to incorporate the balancing strategy of the TSO (see, e.g., \cite{haaberg2016classification}). For example, the TSO can decide to activate mFRR capacity proactively when it expects an increase in system imbalance and the mFRR bids are significantly cheaper \cite{allard2024forecast}.

The imbalance price is calculated ex post, once the information from all optimization cycles within the imbalance settlement period is available. During the settlement period, an estimate of the price can be derived using the information available at each timestep. In Belgium, the exact price formula changed multiple times over the past years. However, in general, the imbalance price can be characterized as a function \cite{bruninx2026herd}
\begin{equation}
    \lambda^{\mathrm{B}} = f\left(\lambda^{\mathrm{aFRR}}, \lambda^{\mathrm{mFRR}}, \lambda^{\mathrm{spot}}\right),
\end{equation}
where $\lambda^{\mathrm{aFRR}}$ denotes the price component arising from aFRR activations, $\lambda^{\mathrm{mFRR}}$ the component arising from mFRR activations and $\lambda^{\mathrm{spot}}$ the component reflecting the spot price. The price function used in this paper is the Belgian imbalance price before the launch of PICASSO and MARI \cite{elia2024brp}, with its details outlined in \ref{app:imbalance_price_formula}.

\subsection{Formulation as a Markov decision process}

To facilitate the application of RL methods, the implicit balancing problem must first be formulated as a Markov Decision Process (MDP) using a tuple $\left(\mathcal{S},\mathcal{A},\mathcal{P}, \mathcal{R}, \gamma\right)$. The state and action spaces are denoted by $\mathcal{S}$ and $\mathcal{A}$, respectively. The reward function is defined as $\mathcal{R}: \mathcal{S} \times \mathcal{A} \rightarrow \mathbb{R}$ and the transition function as $\mathcal{P}: \mathcal{S} \times \mathcal{A} \times \mathcal{S} \rightarrow \left[0,1\right]$. Last, $\gamma \in \left(0,1\right]$ denotes the discount factor for future rewards.

For each timestep $t$, the state space $\mathcal{S}$ combines information related to the wind farm:
\begin{itemize}
    \item The possible power production at the previous timestep $\overline{P}_{t-1}$,
    \item The actual power production at the previous timestep $P_{t-1}$,
    \item The trading schedule for the current timestep $v_t$,
    \item The accumulated position during the settlement period up to the current timestep $\Delta E_t = \sum_{i\in ISP, i<t} (P_i - v_i)\>\Delta t$.
\end{itemize}
with information derived from the minute-level publication by the TSO at that timestep:
\begin{itemize}
    \item The imbalance price $\hat{\lambda}^\B_t$,
    \item The marginal incremental and decremental price,\footnote{Elia defines these values based on the prices of the incremental and decremental bids activated during the settlement period. For the detailed calculation, cf., (\ref{eq:mip}) and (\ref{eq:mdp}).} respectively, $\hat{\lambda}^\mathrm{FRR^+}_t$ and $\hat{\lambda}^\mathrm{FRR^-}_t$,
    \item A forecast of the system imbalance during the imbalance settlement period $\hat{\mathrm{SI}}^\mathrm{ISP}_t$,
    \item The current minute of the settlement period $t^\mathrm{ISP}\in[0,14]$.
\end{itemize}
Note that we include a forecast of the system imbalance to deal with the partial observability of the MDP.

The action is defined as the difference between the power setpoint $x_t$ and the trading schedule $v_t$:
\begin{equation}
    a_t = x_t -v_t \in \mathcal{A},
\end{equation}
where $\mathcal{A}\subset\mathbb{R}$ constitutes a continuous action space.

At the end of each settlement period, a reward is obtained, given by:
\begin{equation}
    r_t = \lambda^{\mathrm{B}}\sum_{i\in ISP} (P_i - v_i)\>\Delta t.
\end{equation}

To transition to the next state, a simulator has been developed that captures the dynamics of both the wind farm and the reserve market. First, the power setpoint $x_t$ is applied to the wind farm, accounting for the possible power and other constraints on the setpoint, which yields the actual power production $P_t$. The residual system imbalance $\mathrm{SI}_t$ is then computed based on the historical system imbalance together with the position of the wind farm and is used to determine which reserves must be activated to resolve the imbalance. The effect of the balancing strategy pursued by the TSO is incorporated using historical data on proactive reserve actions.

Unlike batteries participating in implicit balancing (see, e.g., \cite{rasic2025safe, madahi2025model, madahi2024distributional}), the power production of a wind farm cannot be stored for later use. Therefore, the episode length is set equal to the settlement period (15 minutes). The discount factor $\gamma$ is set to 1, since each minute of the settlement period contributes equally to the final reward.

\section{Soft actor-critic}
\label{sec:sac}

Soft actor-critic \cite{haarnoja2018soft} is an off-policy deep RL algorithm built around two neural networks: an actor $\pi_\phi$, which maps states to a distribution over actions, and a critic $Q_\theta$, which estimates the expected return of state-action pairs. Two design choices distinguish SAC from other actor-critic methods. First, exploration is encouraged through an entropy term in the objective, contrary to TD3 \cite{fujimoto2018addressing}, which injects noise directly into selected actions. Second, training is stabilized through twin critic networks together with target networks $Q_{\theta'}$, where the latter are updated via Polyak averaging rather than direct copying.

At each update, multiple transitions $(s,a,r,s^{\prime})$ are drawn from the replay buffer~$\mathcal{D}$. The critic is then updated by minimizing the soft Bellman residual
\begin{equation}
    J_Q(\theta) = \mathbb{E}_{(s,a)\sim\mathcal{D}}\left[\left(y - Q_\theta\left(s,a\right)\right)^2\right],
\end{equation}
with target
\begin{equation}
    y = r +\gamma \> \mathbb{E}_{s^{\prime} \sim \mathcal{D},a^{\prime}\sim\pi_\phi}\left[Q_{\theta^{\prime}}\left(s^{\prime},a^{\prime}\right) - \alpha \ln \pi_\phi\left(a^{\prime}|s^{\prime}\right)\right],
\end{equation}
where $\alpha$ denotes the temperature parameter, which controls the relative weight of the entropy term, and $a^{\prime}$ the action taken in the next state $s^{\prime}$. The actor, in turn, is learned through the loss function
\begin{equation}
    J_\pi(\phi) = \mathbb{E}_{s \sim \mathcal{D}, a \sim \pi_\phi}[\alpha \> \ln \pi_\phi(a|s) - Q_\theta(s,a)],\label{eq:sac_actor_loss}
\end{equation}
which jointly maximizes the critic's value estimate and the entropy of the policy. Instead of treating $\alpha$ as a fixed hyperparameter, it is adjusted automatically over the course of training, following the procedure introduced in~\cite{haarnoja2018soft_appl}, by minimizing
\begin{equation}
    J(\alpha) = \mathbb{E}_{a \sim \pi_\phi}\left[-\alpha \> \left(\ln \pi_\phi(a|s) +\mathcal{H}\right)\right],
\end{equation}
with $\mathcal{H}$ specifying the target entropy that the policy should attain.

\section{Distributional soft actor-critic}
\label{sec:distsac}


In distributional reinforcement learning \cite{bellemare2017distributional}, the entire distribution of the returns $Z(s,a)$ is considered rather than only the expected value. Translating this to soft actor-critic, the critic network is modified to estimate the return distribution rather than the Q-value. Unlike \cite{duan2021distributional}, which models this distribution using a Gaussian, we follow \cite{dabney2018distributional, ma2025dsac} and use quantile regression, allowing us to also model skewed or heavy-tailed distributions. Specifically, we approximate the return distribution as
\begin{equation}
    Z_\theta(s,a) = \frac{1}{N}\sum_{i=1}^N \delta_{q^{\tau_i}_\theta\left(s,a\right)},
\end{equation}
where $\delta_{z}$ denotes the Dirac function at $z\in\mathbb{R}$, $q^{\tau_i}_\theta\left(s,a\right)$ equals the $\tau_i$-th quantile of the return distribution $Z_\theta$ for a given state-action pair, and $N$ is the number of quantiles.

The loss function of the critic is now adapted to
\begin{equation}
    J_Z(\theta) = \mathbb{E}_{(s,a)\sim\mathcal{D}}\left[ \frac{1}{N}\sum_{j=1}^{N}\sum_{i=1}^{N} \rho^\kappa_{\tau_i}\left(y_{\tau_j} - q^{\tau_i}_\theta\left(s,a\right)\right) \right],
\end{equation}
where $\rho^\kappa_\tau(u)$ is the quantile Huber loss, as defined in \cite{huber1992robust, dabney2018distributional}, and the target
\begin{equation}
    y_{\tau_j} = r +\gamma \mathbb{E}_{s^{\prime} \sim \mathcal{D}, a^{\prime}\sim\pi_\phi}\left[q^{\tau_j}_{\theta^{\prime}}\left(s^{\prime},a^{\prime}\right) - \alpha \ln \pi_\phi\left(a^{\prime}|s^{\prime}\right)\right].
\end{equation}
The risk-neutral actor can be trained by minimizing (\ref{eq:sac_actor_loss}) and taking into account that
\begin{equation}
    Q_\theta(s,a) = \mathbb{E}\left[Z_\theta\left(s,a\right)\right].
\end{equation}
To learn a risk-averse strategy, the actor loss should be modified to
\begin{equation}
    J_\pi(\phi) = \mathbb{E}_{s \sim \mathcal{D}, a \sim \pi_\phi}\left[\alpha \ln \pi_\phi(a|s) - \Psi\left[Z_\theta(s,a)\right]\right],\label{eq:sac_actor_loss_risk}
\end{equation}
where $\Psi\left(\cdot\right)$ is a risk measure. In this work, we will consider the conditional value at risk (CVaR):
\begin{equation}
    \mathrm{CVaR}_\beta(X) = \frac{1}{\beta}\int_{0}^{\beta} F^{-1}_X(u) \, du,
\end{equation}
with $\beta$ the risk level, $X$ a random variable and $F^{-1}_X(u)$ the quantile function, which is the inverse of the cumulative distribution function. Unlike value at risk, conditional value at risk satisfies the subadditivity property of a coherent risk measure~\cite{artzner1999coherent}, making it a preferred metric in risk management.

\section{Numerical results}
\label{sec:results}

\subsection{Set-up and training convergence}

\begin{table}[t!]
    \centering
    \caption{Overview of the control methods considered in this study.}
    \label{table:model_overview}
    \resizebox{0.85\linewidth}{!}{
    \begin{tabularx}{\textwidth}{l X}
        \toprule
        Model type & Description \\
        \midrule
        No control  &  Baseline strategy where the wind farm always produces at possible power. \\
        Price threshold   &  Baseline strategy, often used in industry, where the setpoints are determined using the minute-level imbalance prices published by the TSO. The details of this strategy are described in \ref{app:price_threshold}. \\
        SAC     &  RL-based control strategy where the setpoints are determined using soft actor-critic. \\
        distSAC           &  RL-based control strategy where the setpoints are determined using distributional soft actor-critic with the objective to maximize expected returns. \\
        distSAC $\mathrm{CVaR}_\beta$  &  RL-based control strategy where the setpoints are determined using distributional soft actor-critic with the objective to minimize risk defined using conditional value at risk with risk-level $\beta$.\\
        \bottomrule
    \end{tabularx}
    }
\end{table}

Experiments are conducted using Belgian data from 2022 and 2023. The first year and a half is used to train the RL agents and the last six months for evaluation. The data comes from the Open Data Platform by Elia~\cite{eliaopendata} and the SCADA system of Norther, a Belgian wind farm with an installed capacity of 370 MW. The farm-level possible power is defined using the wind speed measurements per turbine combined with the power curve of the manufacturer. The trading schedule is set equal to the day-ahead power forecast, with forecasts obtained using the approach described in \cite{bruninx2027probabilistic}. The system imbalance forecasts are obtained using Chronos-2 \cite{ansari2025chronos} with the published minute-level system imbalances. These forecasts are updated every minute based on the most recent information. All other variables can be derived from the data directly. 

Table~\ref{table:model_overview} provides an overview of all control methods considered in this study. Each method is evaluated with a setpoint resolution of both 1 minute and 15 minutes, where the 15-minute controllers apply a fixed setpoint at the start of each settlement period. For SAC and distSAC, hyperparameter tuning was performed using the last three months of the training period as a validation set, after which the final agents were trained on the full training period. The actor networks use a learning rate of \num{3e-4}, while the critic networks and the optimizer of the temperature parameter $\alpha$ use a learning rate of \num{1e-3}. All networks are implemented as multi-layer perceptrons with two hidden layers, using $256$ neurons per layer for the 1-minute controllers and $32$ for the 15-minute controllers. The soft update factor $\tau$ is set to $0.01$, the replay buffer size to \num{1e7}, and the batch size to $1024$ for SAC and $256$ for distSAC. For distSAC, we use $32$ quantiles and set the parameter $\kappa$ of the quantile Huber loss to $1$. 

\begin{figure}[t!]
    \centering
    \includegraphics[width=0.75\linewidth]{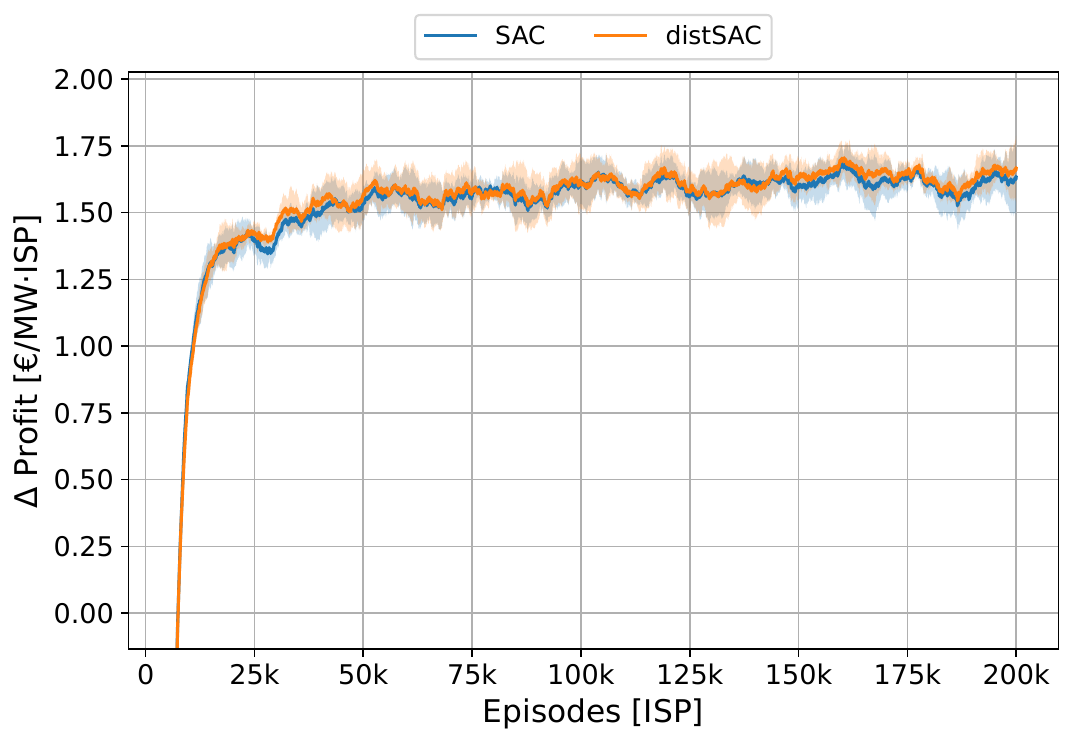}
    \caption{Training convergence of the minute-level RL agents. The x-axis denotes the number of episodes, the y-axis the moving average of the additional profit compared to the price threshold controller. The solid line shows the mean and the shaded area the standard deviation across the five seeds.}
    \label{fig:training_convergence}
\end{figure}

The RL agents are trained for $200$k episodes using five different random seeds. Figure~\ref{fig:training_convergence} shows the training convergence of the risk-neutral agents with 1-minute setpoints. We visualize convergence as the additional profit relative to the price threshold controller, which cancels out episodes in which no controller can reduce imbalance costs. This can occur when the power forecast overestimates the actual production and the settlement price is positive. The solid lines denote the mean over the five seeds and the shaded areas the standard deviation across seeds. Both agents outperform the price threshold controller after approximately $8$k episodes and improve rapidly during the first $25$k episodes. From around $50$k episodes onward, the additional profit stabilizes between $1.50$ and $1.75$~€ per MW installed per settlement period. Throughout training, distSAC attains marginally higher profits than SAC, although the difference is small compared to the variability across seeds. The risk-averse agents are obtained by finetuning the risk-neutral distSAC agent for $50$k episodes, optimizing conditional value at risk with risk levels $\beta \in \{0.01, 0.05\}$.


\subsection{Out-of-sample performance}


\begin{table}[b!]
\caption{Summary statistics over the evaluation period of the imbalance costs ($+$) or revenues ($-$) for the different policies, in € per MW installed per settlement period. Mean denotes the average over all settlement periods, and P05 and P95 the 5\% and 95\% quantiles, respectively. For the RL policies, each statistic is reported as the average $\pm$ standard deviation over five training seeds, the baselines are deterministic and therefore have no standard deviation.}
\label{table:model_pnl}
\centering
\resizebox{0.8\linewidth}{!}{
\begin{tabular}{llccc}
\toprule
  &  & \phantom{$-$}Mean\phantom{\,{\scriptsize$\pm$0.00}} & \phantom{$-$}P05\phantom{\,{\scriptsize$\pm$0.00}} & P95\phantom{\,{\scriptsize$\pm$0.00}} \\
\toprule
\noalign{\vskip 0.5ex}
 & No control & \phantom{$-$}3.22\phantom{\,{\scriptsize$\pm$0.00}} & \phantom{0}$-$6.55\phantom{\,{\scriptsize$\pm$0.00}} & 22.13\phantom{\,{\scriptsize$\pm$0.00}} \\
\noalign{\vskip 0.5ex}
\cline{1-5}
\noalign{\vskip 0.75ex}
\multirow[t]{3}{*}{\fifteenmin} & Price threshold & \phantom{$-$}2.57\phantom{\,{\scriptsize$\pm$0.00}} & \phantom{0}$-$4.33\phantom{\,{\scriptsize$\pm$0.00}} & 19.04\phantom{\,{\scriptsize$\pm$0.00}} \\
 & SAC & \phantom{$-$}1.81\,{\scriptsize$\pm$0.06} & \phantom{0}$-$8.27\,{\scriptsize$\pm$0.24} & 17.19\,{\scriptsize$\pm$0.40} \\
 & distSAC & \phantom{$-$}1.78\,{\scriptsize$\pm$0.05} & \phantom{0}$-$8.16\,{\scriptsize$\pm$0.24} & 17.14\,{\scriptsize$\pm$0.25} \\
\noalign{\vskip 0.75ex}
\cline{1-5}
\noalign{\vskip 0.75ex}
\multirow[t]{3}{*}{\onemin} & Price threshold & \phantom{$-$}1.47\phantom{\,{\scriptsize$\pm$0.00}} & \phantom{0}$-$8.39\phantom{\,{\scriptsize$\pm$0.00}} & 16.30\phantom{\,{\scriptsize$\pm$0.00}} \\
 & SAC & $-$0.02\,{\scriptsize$\pm$0.09} & $-$15.52\,{\scriptsize$\pm$0.15} & 15.36\,{\scriptsize$\pm$0.50} \\
 & distSAC & $-$0.12\,{\scriptsize$\pm$0.03} & $-$15.49\,{\scriptsize$\pm$0.20} & 15.10\,{\scriptsize$\pm$0.02} \\
\bottomrule
\end{tabular}
}
\end{table}

We now study the imbalance cost savings achieved by the different controllers for 1-minute and 15-minute setpoint resolutions. For this analysis, we do not consider the risk-averse distSAC agents. Table~\ref{table:model_pnl} reports summary statistics of the imbalance costs over the evaluation period, where for the RL policies we report the average and standard deviation over five training seeds. First, the empirical results show that each control policy with a 1-minute setpoint resolution yields a substantial improvement over the policy obtained using the same algorithm with a 15-minute resolution. This illustrates the value of receiving additional information within the settlement period and the ability to act on it. At the 1-minute resolution, the price threshold controller reduces the imbalance cost by $54\%$ relative to the no-control baseline, whereas SAC and distSAC reduce it by $101\%$ and $104\%$, respectively, thereby turning the imbalance cost into a small revenue. For distSAC, this holds for each of the five training seeds, whereas for SAC it holds for four of the five seeds, with the remaining seed still incurring a small cost. At the 15-minute resolution, by contrast, all controllers still incur an imbalance cost, with reductions of $20\%$ for the price threshold controller and $44\%$ and $45\%$ for SAC and distSAC. Last, when comparing the SAC and distSAC policies, we find that distSAC attains slightly lower imbalance costs than SAC at the 1-minute resolution. Although this difference is small, it is consistent across training seeds, as even the least profitable distSAC seed outperforms the most profitable SAC seed. At the 15-minute resolution, by contrast, the two policies perform nearly identically.

\begin{figure}[p!]
    \centering

    \begin{subfigure}[b]{\textwidth}
        \centering
        \includegraphics[width=0.48\textwidth]{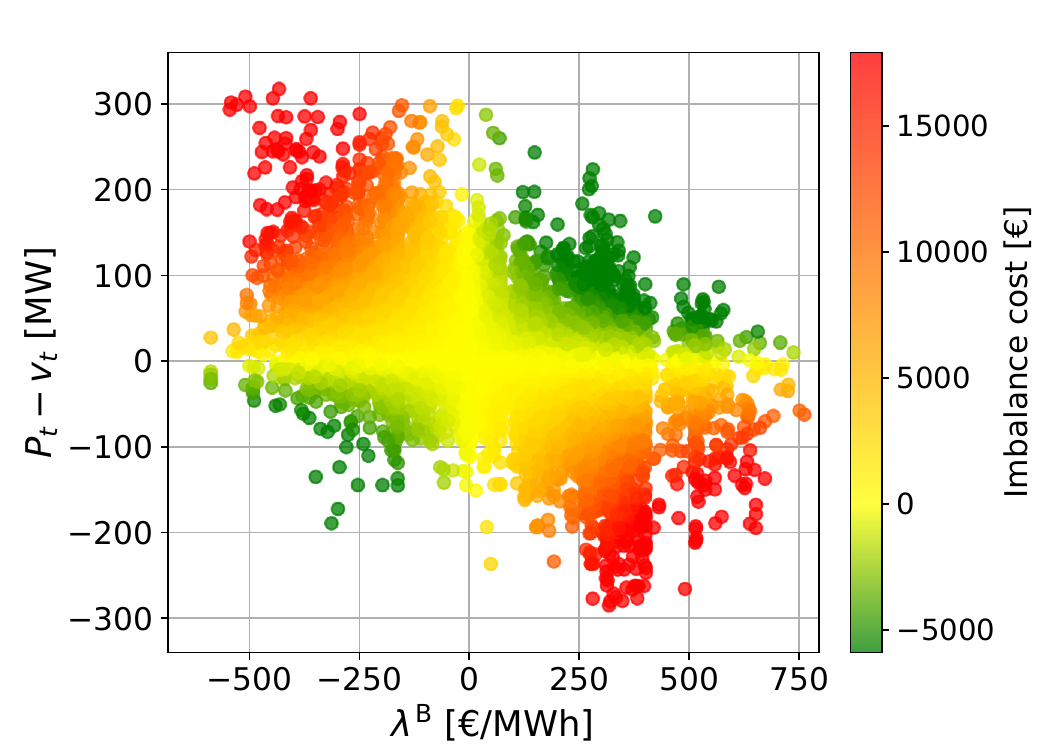}
        \hfill
        \includegraphics[width=0.48\textwidth]{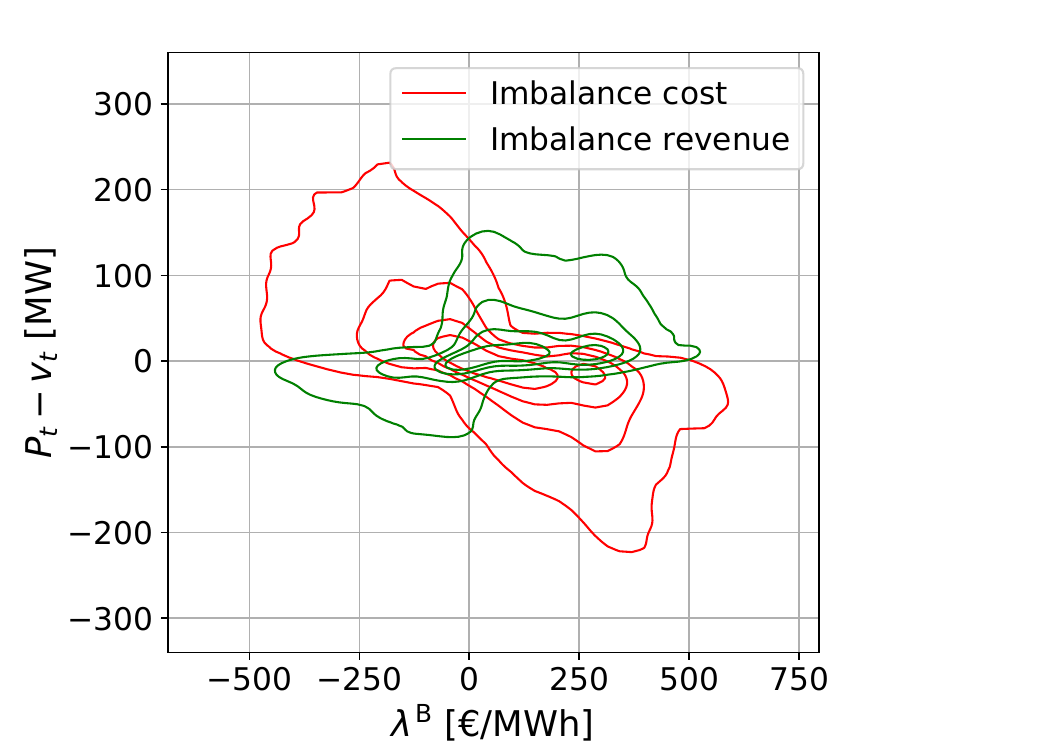}
        \caption{No control}
        \label{fig:no_curtailment}
    \end{subfigure}

    \vspace{1em}

    \begin{subfigure}[b]{\textwidth}
        \centering
        \includegraphics[width=0.48\textwidth]{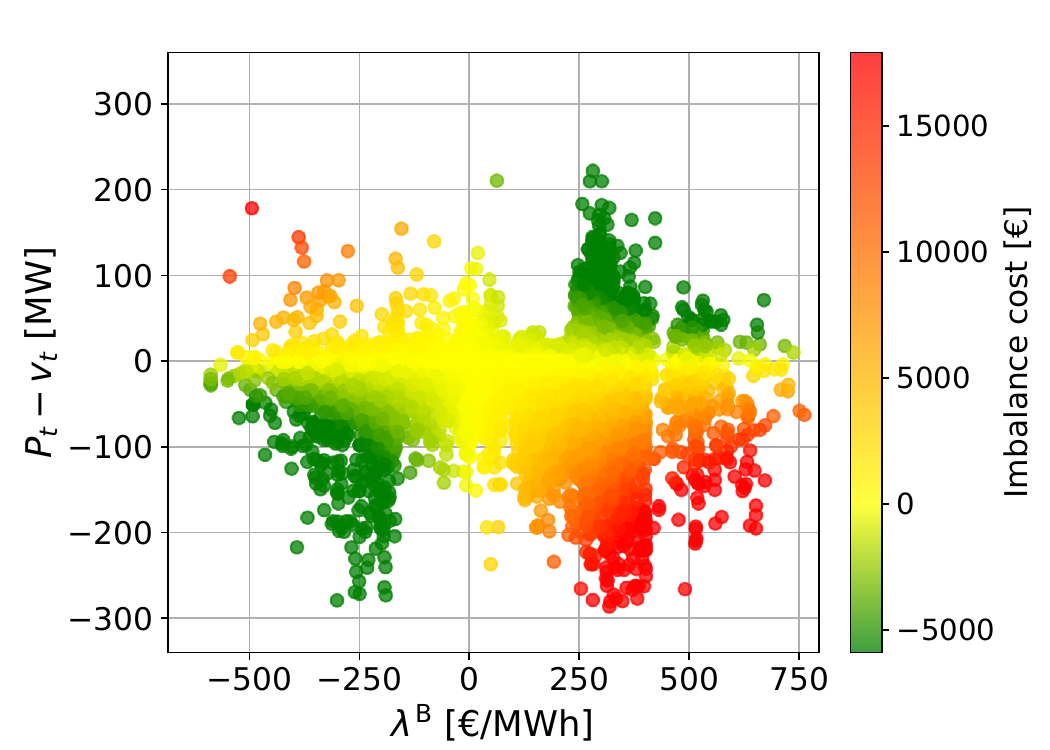}
        \hfill
        \includegraphics[width=0.48\textwidth]{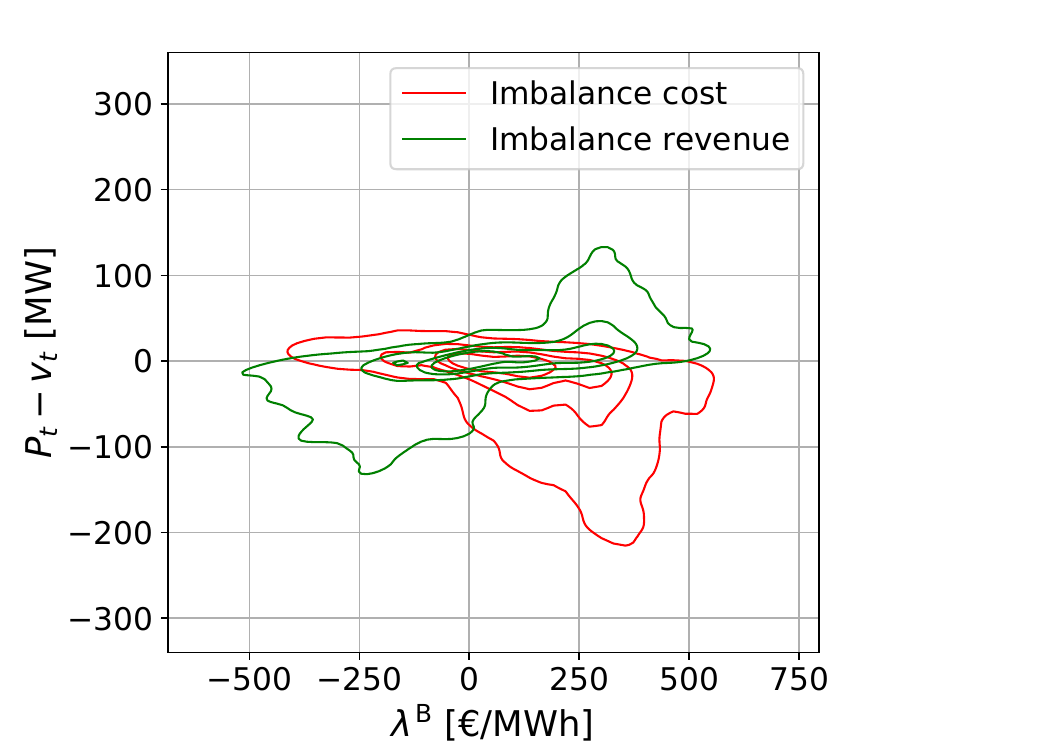}
        \caption{Price threshold}
        \label{fig:price_threshold}
    \end{subfigure}

    \vspace{1em}

    \begin{subfigure}[b]{\textwidth}
        \centering
        \includegraphics[width=0.48\textwidth]{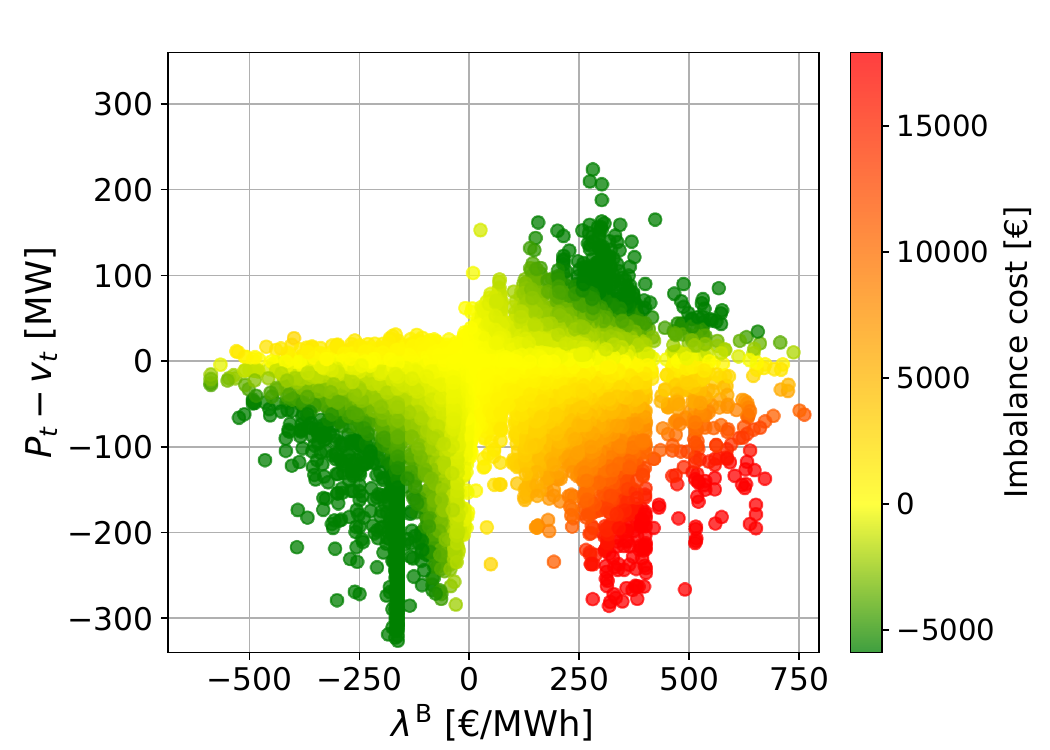}
        \hfill
        \includegraphics[width=0.48\textwidth]{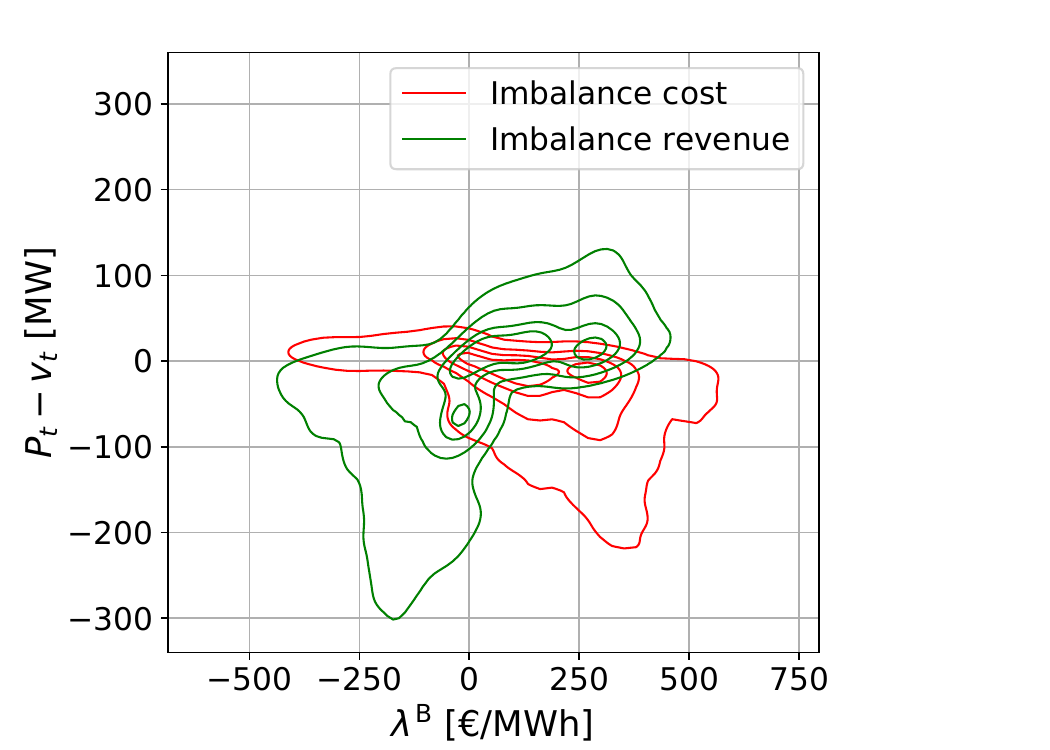}
        \caption{distSAC}
        \label{fig:distsac}
    \end{subfigure}

    \caption{Scatter (left) and contour (right) plots visualizing the impact on imbalance costs/revenues of the average open position over the settlement period (y-axis) and the settlement price (x-axis). Both controllers use 1-minute setpoints. For distSAC, a single representative training seed is shown, namely the one whose imbalance cost over the evaluation period is the median of the five seeds.}
    \label{fig:model_comparison}
\end{figure}

To further illustrate the added value of real-time asset control during imbalance settlement, Figure \ref{fig:model_comparison} shows the impact on imbalance costs for different control policies. The impact is visualized as a function of the open position over the settlement period and the corresponding settlement price. In Figure \ref{fig:no_curtailment}, the imbalance costs without control actions are shown. These costs result from the forecast errors in the day-ahead stage and can be partly reduced with real-time control. In particular, the observations in the top-left quadrant, where the wind farm has a surplus while the settlement price is negative, represent avoidable costs, since the wind farm could have curtailed its power production. In Figures \ref{fig:price_threshold} and \ref{fig:distsac}, the behavior of, respectively, the price threshold and distSAC control policy is shown. We observe that both control policies are capable of reducing the costs incurred in the top-left quadrant. That we cannot reduce the costs in the bottom-right quadrant is expected, since in this quadrant the system is facing a deficit and the wind farm can only produce less than its trading schedule. Furthermore, we find that not all costs in the top-left quadrant could be avoided and that the observations in the top-right quadrant are also affected by the controllers. This is because the settlement price is only known after the settlement period and the controllers are therefore making decisions under uncertainty. When comparing the contour plots of the two control methods, we observe a significant difference in the distribution of the observations. Compared to the distSAC policy, the probability density of the price threshold controller is more concentrated around zero on the y-axis, which corresponds to a power production approximately equal to the trading schedule. In contrast, the density of the distSAC policy is spread more widely over the bottom-left and top-right quadrants, where curtailment actions can yield imbalance revenue. 

\subsection{Risk management and price impact}


In this section, we want to assess whether the distributional SAC agents can be adapted to pursue a risk-averse objective. To this end we define two risk-averse policies with 1-minute setpoints which optimize conditional value at risk with a risk level $\beta$ of $0.05$ and $0.01$. The summary statistics for the two policies over the evaluation period are given in Table \ref{table:model_pnl_risk}. Based on the works of~\cite{madahi2024distributional,madahi2025risk}, we expect that the risk-averse policies would reduce the tail risk of the control strategy at the expense of lower overall savings of the imbalance costs. However, in our experiments, we find that all policies attain a similar imbalance cost reduction and that there are no significant differences in the aggregate risk measures in Table~\ref{table:model_pnl_risk}. This difference could be explained by the fact that curtailment mainly serves to reduce exposure to imbalance costs, whereas a battery can also arbitrage between low- and high-price periods. The main risk of curtailment is therefore a positive settlement price, caused either by the price impact of the control strategy or by an incorrect anticipation of the settlement price. Furthermore, the summary statistics in Table~\ref{table:model_pnl_risk} consider the profit and loss during imbalance settlement, which also includes the effect of day-ahead forecast errors. As shown in Figure~\ref{fig:model_comparison}, the potential imbalance cost reduction of the controller depends on these forecasts. Hence, a more insightful analysis would distinguish between the risk originating from the forecast errors and the risk introduced by the control actions.

\begin{table}[b!]
\caption{Summary statistics over the evaluation period of the imbalance costs (+) or revenues ($-$) for the risk-averse policies trained with risk levels $\beta \in \{0.01, 0.05\}$, in € per MW installed per settlement period. Mean denotes the average over all settlement periods, and P90 and P95 the 90\% and 95\% quantiles, respectively. The CVaR risk levels are defined with respect to the profit distribution, consistent with the objective of the actor. Each statistic is reported as the average $\pm$ standard deviation over five training seeds. The risk-neutral distSAC policy is included for reference.}
\label{table:model_pnl_risk}
\centering
\resizebox{0.85\linewidth}{!}{
\begin{tabular}{llccccc}
\toprule
 & $\beta$ & \phantom{$-$}Mean\phantom{\,{\scriptsize$\pm$0.00}} & P90\phantom{\,{\scriptsize$\pm$0.00}} & P95\phantom{\,{\scriptsize$\pm$0.00}} & $\mathrm{CVaR}_{0.05}$ & $\mathrm{CVaR}_{0.01}$ \\
\toprule
\noalign{\vskip 0.5ex}
\multirow[t]{2}{*}{distSAC (CVaR)} & 0.01 & $-$0.07\,{\scriptsize$\pm$0.06} & 7.34\,{\scriptsize$\pm$0.02} & 15.09\,{\scriptsize$\pm$0.01} & 29.34\,{\scriptsize$\pm$0.06} & 54.91\,{\scriptsize$\pm$0.15} \\
 & 0.05 & $-$0.09\,{\scriptsize$\pm$0.04} & 7.37\,{\scriptsize$\pm$0.06} & 15.12\,{\scriptsize$\pm$0.05} & 29.39\,{\scriptsize$\pm$0.12} & 54.86\,{\scriptsize$\pm$0.12} \\
\noalign{\vskip 0.75ex}
\cline{1-7}
\noalign{\vskip 0.75ex}
distSAC & -- & $-$0.12\,{\scriptsize$\pm$0.03} & 7.39\,{\scriptsize$\pm$0.05} & 15.10\,{\scriptsize$\pm$0.02} & 29.36\,{\scriptsize$\pm$0.05} & 54.82\,{\scriptsize$\pm$0.06} \\
\bottomrule
\end{tabular}
}
\end{table}

\begin{figure}[h!]
    \centering
    \begin{subfigure}[b]{0.48\linewidth}
        \centering
        \includegraphics[width=\linewidth]{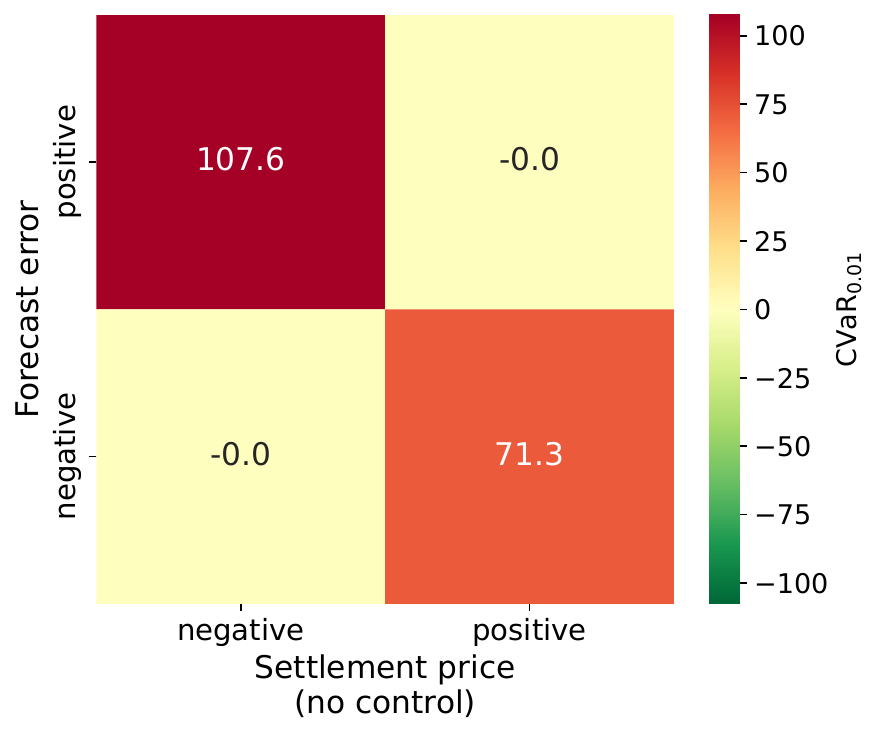}
        \caption{No control}
        \label{fig:risk_cvar_no_curtailment}
    \end{subfigure}
    \hfill
    \begin{subfigure}[b]{0.48\linewidth}
        \centering
        \includegraphics[width=\linewidth]{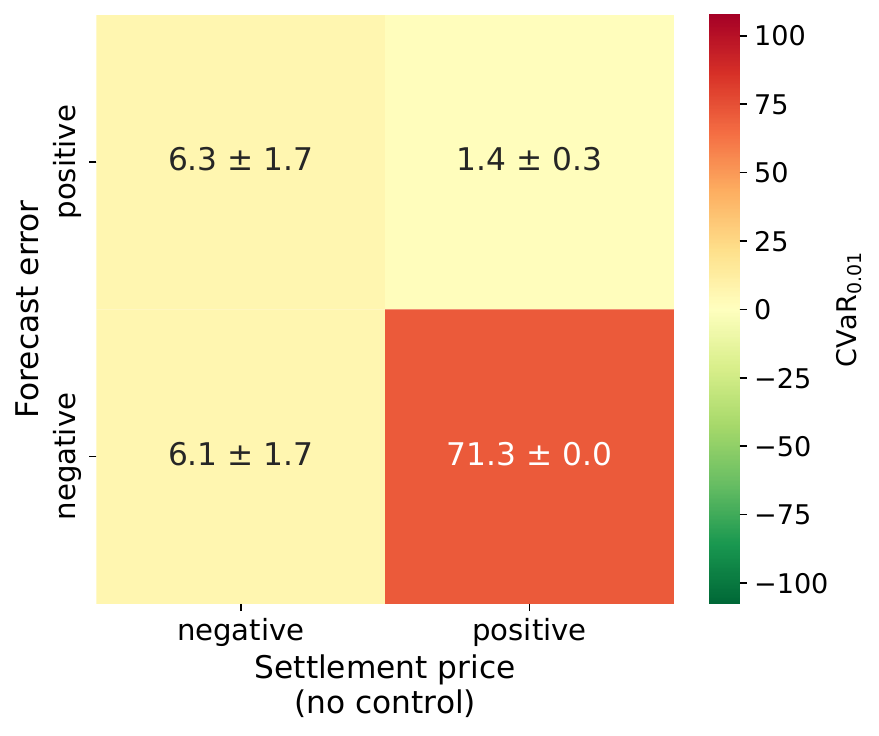}
        \caption{distSAC}
        \label{fig:risk_cvar_distsac}
    \end{subfigure}
    \\[1ex]
    \begin{subfigure}[b]{0.48\linewidth}
        \centering
        \includegraphics[width=\linewidth]{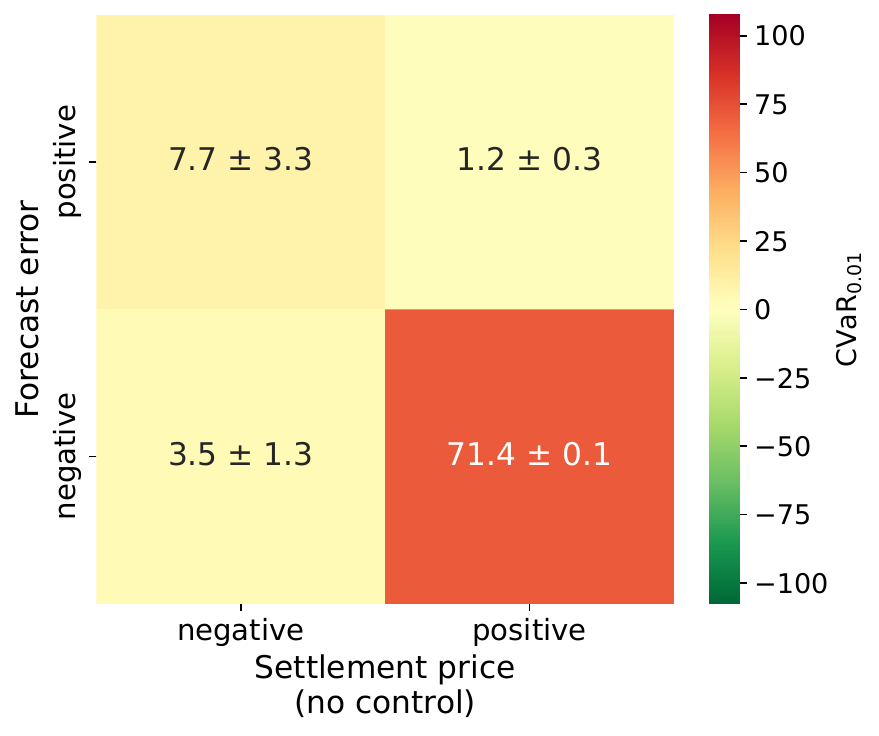}
        \caption{distSAC $\mathrm{CVaR}_{0.05}$}
        \label{fig:risk_cvar_beta05}
    \end{subfigure}
    \hfill
    \begin{subfigure}[b]{0.48\linewidth}
        \centering
        \includegraphics[width=\linewidth]{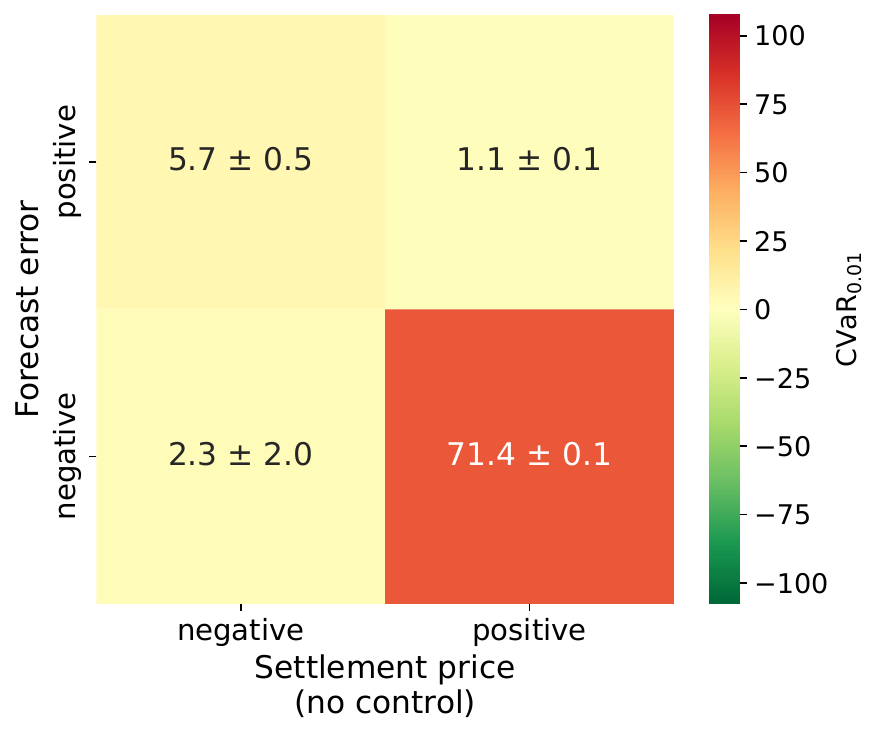}
        \caption{distSAC $\mathrm{CVaR}_{0.01}$}
        \label{fig:risk_cvar_beta01}
    \end{subfigure}
    \caption{Out-of-sample conditional value at risk ($\beta=0.01$) for the no-control baseline and the distSAC policies. Positive values represent imbalance costs, but the risk level is defined with respect to the profit distribution to remain consistent with the reward definition.  The results are grouped by the forecast error, i.e., the average difference between the possible power and the trading schedule during the settlement period, and the settlement price without control actions. For the distSAC policies, each statistic is reported as the average $\pm$ standard deviation over five training seeds, the no-control baseline is deterministic and therefore has no standard deviation.}
    \label{fig:risk_per_scenario}
\end{figure}

To make this distinction, Figure~\ref{fig:risk_per_scenario} visualizes the conditional value at risk with a risk level $\beta$ of $0.01$, grouped by the forecast error and the settlement price without control actions. The results for $\beta=0.05$ can be found in~\ref{app:add_exp}. Since the scenarios are independent of the control strategy, each subfigure contains the same settlement periods per quadrant, such that differences between the subfigures isolate the effect of the control policy. In Figure~\ref{fig:risk_cvar_no_curtailment}, the imbalance risk is visualized for the baseline strategy without control. The matrix shows how the tail risk is distributed between the scenario with a positive forecast error and a negative settlement price and the scenario with the positive price and a negative forecast error. The other two scenarios do not represent any risk since they result in imbalance revenue. Figure~\ref{fig:risk_cvar_distsac} shows how the risk-neutral distSAC agent is able to reduce the risk in the top-left quadrant, but at the cost of increasing the risk in the top-right and bottom-left quadrants. In the top-left quadrant, the remaining tail risk can have two causes: either the wind farm did not curtail because it did not anticipate a negative settlement price, or it did curtail, but this turned the settlement price positive, i.e., an overshoot. The increase in risk in the bottom-left quadrant is also caused by overshoots. Here, the settlement price would have been negative without control, but additional curtailment turned it positive. In the top-right quadrant, by contrast, the increase stems from curtailment in anticipation of a negative price, while the settlement price was positive regardless of the control actions. Last, in the bottom-right quadrant, the wind farm has a deficit while the settlement price is positive. Since curtailment can only increase this deficit, the best action is to produce at possible power and no control policy can reduce the associated risk. Figures~\ref{fig:risk_cvar_beta05}~and~\ref{fig:risk_cvar_beta01} show the tail risk matrix for the risk-averse control strategies. Both risk-averse policies slightly reduce the tail risk in the top-right and bottom-left quadrants, whereas the effect in the top-left quadrant is mixed. One possible explanation is that the risk-neutral policy already substantially reduces the risk of the top-left quadrant compared to the no-control baseline, which may leave limited room for further improvement in terms of aggregated risk.

\begin{figure}[t!]
    \centering
    \includegraphics[width=0.6\linewidth]{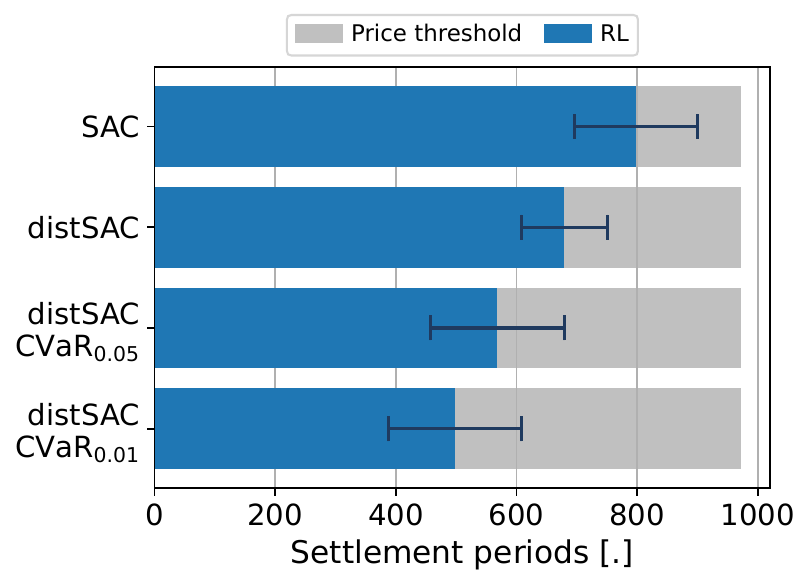}
    \caption{Number of settlement periods with an overshoot for each control policy. Overshoots occur when the settlement price would have remained negative without curtailment, but the curtailment actions resulted in a positive price. The blue bars denote the average number of overshoots over the five training seeds and the error bars one standard deviation across seeds. The grey bar indicates the number of overshoots for the deterministic price threshold controller.}
    \label{fig:price_impact}
\end{figure}

Nonetheless, the risk-averse objective does change the learned policy. On average, the time spent in curtailed operation decreases by $11\%$ for $\beta=0.05$ and by $15\%$ for $\beta=0.01$ compared to risk-neutral distSAC. This reduction is observed in the majority of the runs, with four of the five seeds showing less time in curtailed operation for both values of $\beta$. The decrease in curtailments helps to avoid overshoots, as shown in Figure~\ref{fig:price_impact}. On average, the number of overshoots decreases by $16\%$ for $\beta=0.05$ and by $27\%$ for $\beta=0.01$ compared to risk-neutral distSAC. The size of this effect, however, varies considerably across seeds. Since each risk-averse policy is finetuned from the risk-neutral policy of the same seed, this variability is partly inherited from the starting point. Seeds whose risk-neutral policy overshot most often show the largest reductions (up to $50\%$), whereas seeds with fewer overshoots change only marginally. To illustrate this, we include the per-seed change relative to the risk-neutral policy in \ref{app:add_exp}.

\section{Discussion and conclusion}
\label{sec:discussion}

In this work we studied how wind farms can participate in implicit balancing to reduce their imbalance costs. Given the uncertainty of the settlement price and the impact the control strategy can have on this price, we adopted distSAC, a method from the field of distributional reinforcement learning. An advantage of this method compared to SAC is that it can also learn risk-averse strategies by modifying the objective of the actor network to include a risk measure. In our work, we considered risk-averse strategies based on conditional value at risk, a coherent risk measure. To capture the price impact of the control strategy, we used the minute-level imbalance price simulator of~\cite{bruninx2026herd}, which replicates the actual imbalance price formation based on historical data provided by the TSO. We combined this simulator with data from the SCADA system of Norther, a Belgian offshore wind farm, to determine the possible power based on actual wind conditions. The findings show that, with 1-minute setpoints, distSAC substantially reduces imbalance costs, outperforming a price threshold controller and slightly outperforming SAC. With the commonly adopted 15-minute setpoints, the cost reductions are considerably smaller for all controllers, and distSAC and SAC perform nearly identically. This highlights the value of acting on the minute-level information published by the TSO. Furthermore, unlike SAC, distSAC allows finetuning the risk-neutral agent using conditional value at risk. This finetuning step was found to yield a control strategy that better accounts for its own impact on the imbalance price, resulting in fewer overshoots, while attaining almost the same degree of imbalance cost savings.

An important direction for future work would be to improve the interpretability of the control method. While deep reinforcement learning methods have shown impressive results over the past decade, their lack of interpretability poses a constraint for deployment on real-world systems \cite{whittlestone2021societal, glanois2024survey}. This is particularly important in our setting, since the asset under control can have a significant impact on the power grid, and incorrect control actions may result in considerable financial losses. One way to achieve interpretability for RL methods is by employing soft decision trees \cite{frosst2017distilling}, which have already shown promising results in the context of home energy management \cite{gokhale2024distill2explain}.

\section*{Acknowledgements}

The authors gratefully acknowledge the financial support of the Flemish Government through the Flanders AI Research Program, the INSPIRE project (project no. SBEP2023-440) funded by the Sustainable Blue Economy Partnership and the Blue Cluster project called CORE funded by Flanders Innovation \& Entrepreneurship (VLAIO).

\section*{CRediT authorship contribution statement}

\textbf{Max Bruninx}: Conceptualization, Formal analysis, Investigation, Methodology, Software, Visualization, Writing - original draft. \textbf{Seyed Soroush Karimi Madahi}: Software, Writing - review \& editing. \textbf{Timothy Verstraeten}: Supervision, Conceptualization, Writing - review \& editing. \textbf{Jan Decuyper}: Conceptualization, Writing - review \& editing. \textbf{Chris Develder}: Writing - review \& editing. \textbf{Ann Nowé}: Funding acquisition. \textbf{Jan Helsen}: Supervision, Funding acquisition.





\bibliographystyle{elsarticle-num}
\FloatBarrier
\bibliography{ref}

\appendix

\section{Imbalance price}
\label{app:imbalance_price_formula}

In this work, we consider the imbalance price formula in place before the launch of PICASSO and MARI \cite{elia2024brp}:
\begin{equation}
    \lambda^\textrm{B}=\begin{cases}
        \max\left(\lambda^\aFRR, \lambda^\mFRR\right)\quad \mathrm{SI}\leq0,\\
        \min\left(\lambda^\aFRR, \lambda^\mFRR\right)\quad \mathrm{SI}>0,\\
    \end{cases}
\end{equation}
where $\mathrm{SI}$ denotes the system imbalance during the ISP. Note that this formula does not include a spot price component. The spot price is, however, introduced in a later version of the price formula through the deadband applied to system imbalances between -25 and 25 MW.

The aFRR component is given by the weighted average price of the activated volumes to counteract the system imbalance:
\begin{equation}
    \lambda^{\aFRR} =
    \begin{cases}
        \dfrac{\sum_{a^{+}\in \aFRR^{+}} \mu^{a^{+}}_t\, b^{a^{+}}_t}{\sum_{a^{+}\in \aFRR^{+}} b^{a^{+}}_t}, & \mathrm{SI}\leq 0\\[3mm]
        \dfrac{\sum_{a^{-}\in \aFRR^{-}} \mu^{a^{-}}_t\, b^{a^{-}}_t}{\sum_{a^{-}\in \aFRR^{-}} b^{a^{-}}_t}, & \mathrm{SI}> 0
    \end{cases}
\end{equation}
In case there are no aFRR activations, this component is set equal to the average of the value of additional activation (VoAA)\footnote{As defined in ``T\&C BRP'' \cite{elia2024brp}: The cheapest energy bid during the imbalance settlement period considering both aFRR and mFRR in the local merit order for a given direction.} in the upward and downward direction.

The mFRR component is given by the marginal mFRR bid that has been activated over the imbalance settlement period:
\begin{equation}
    \lambda^{\mFRR} = \begin{cases}
        \max_{t\in ISP} \lambda^{\mFRR^{+}}_t\quad \mathrm{SI}\leq0\\
        \min_{t\in ISP} \lambda^{\mFRR^{-}}_t\quad \mathrm{SI} >0.
    \end{cases}
\end{equation}
In case there are no mFRR activations, this component does not exist.

To correct for situations with low imbalance prices and large system imbalances, the Belgian TSO employs the alpha component
\begin{equation}
    \alpha^\mathrm{B} = \left(a + \frac{b}{1+e^\frac{c-x}{d}}\right) cp,
\end{equation}
with $a=0,\ b=200,\ c=450,\ d=65,\ x=$ average of the system imbalance during the current ISP and the system imbalance during the previous ISP and 
\begin{align}
    cp =\begin{cases}
        \mathrm{clip}(\frac{400-\lambda^{\B}}{200})& \mathrm{SI}<-150,\\
        -\mathrm{clip}(\frac{\lambda^{\B}+200}{200})& \mathrm{SI}>150,\\
        0& \mathrm{SI} \in [-150, 150],
    \end{cases}
\end{align}
with $\mathrm{clip}(x) = \min\left(1, \max(0,x)\right)$ a function which clips the values between 0 and 1. Afterwards, the imbalance price is adapted using
\begin{equation}
    \lambda^{\B}\leftarrow\lambda^{\B}+\alpha^\mathrm{B}.
\end{equation}

Finally, Elia also publishes the marginal incremental and decremental prices, which are, respectively, defined as
\begin{equation}
    \lambda^{\mathrm{FRR}^+} = \max(\dfrac{\sum_{a^{+}\in \aFRR^{+}} \mu^{a^{+}}_t\, b^{a^{+}}_t}{\sum_{a^{+}\in \aFRR^{+}} b^{a^{+}}_t}; \max_{t\in ISP} \lambda^{\mFRR^{+}}_t),\label{eq:mip}
\end{equation}
\begin{equation}
    \lambda^{\mathrm{FRR}^-} = \min(\dfrac{\sum_{a^{-}\in \aFRR^{-}} \mu^{a^{-}}_t\, b^{a^{-}}_t}{\sum_{a^{-}\in \aFRR^{-}} b^{a^{-}}_t}; \min_{t\in ISP} \lambda^{\mFRR^{-}}_t).\label{eq:mdp}
\end{equation}

\section{Price threshold controller}
\label{app:price_threshold}

The power setpoint $x_t$ chosen by the price threshold controller is determined by the most recently published imbalance price estimate $\hat{\lambda}^{\mathrm{B}}_t$:
\begin{equation}
    x_t = \begin{cases}
        \overline{P}_t\quad\text{ if }\hat{\lambda}^{\mathrm{B}}_t>\overline{\lambda}^\mathrm{B}\\
        \underline{P}\quad\ \text{ if }\hat{\lambda}^{\mathrm{B}}_t<\underline{\lambda}^\mathrm{B}\\
        v_t\ \ \quad\text{else}.\\
    \end{cases}
\end{equation}
The price thresholds $\overline{\lambda}^\mathrm{B}$ and $\underline{\lambda}^\mathrm{B}$ are calibrated from historical price data according to the procedure described in \cite{bruninx2026herd}, yielding $\underline{\lambda}^\mathrm{B}=-186$ €/MWh and $\overline{\lambda}^\mathrm{B}=230$ €/MWh.

\newpage
\section{Additional results}
\label{app:add_exp}

\begin{figure}[h!]
    \centering
    \begin{subfigure}[b]{0.48\linewidth}
        \centering
        \includegraphics[width=\linewidth]{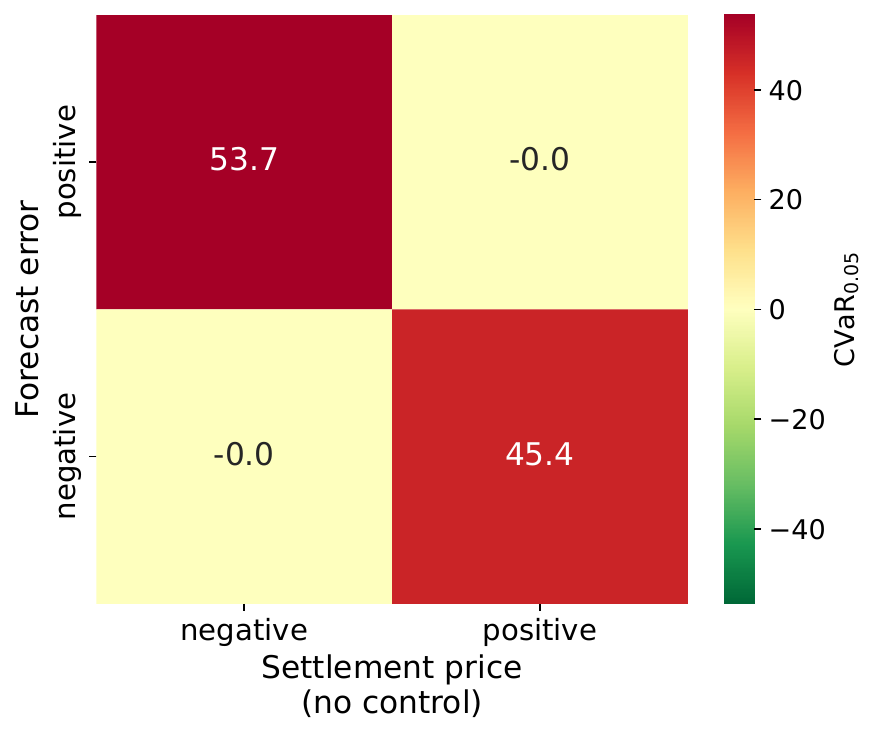}
        \caption{No control}
    \end{subfigure}
    \hfill
    \begin{subfigure}[b]{0.48\linewidth}
        \centering
        \includegraphics[width=\linewidth]{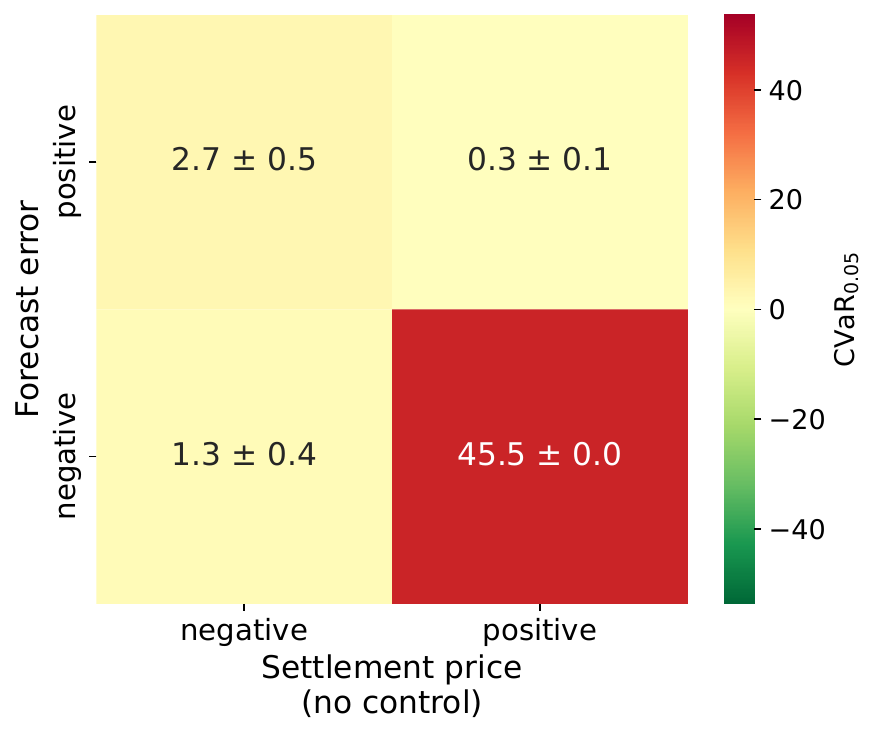}
        \caption{distSAC}
    \end{subfigure}
    \\[1ex]
    \begin{subfigure}[b]{0.48\linewidth}
        \centering
        \includegraphics[width=\linewidth]{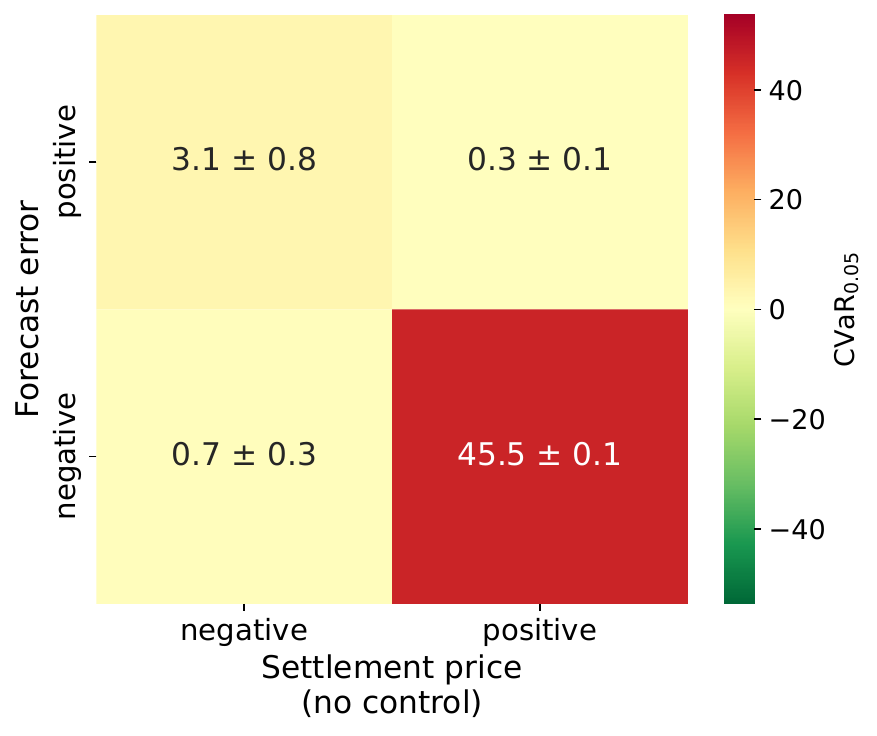}
        \caption{distSAC $\mathrm{CVaR}_{0.05}$}
    \end{subfigure}
    \hfill
    \begin{subfigure}[b]{0.48\linewidth}
        \centering
        \includegraphics[width=\linewidth]{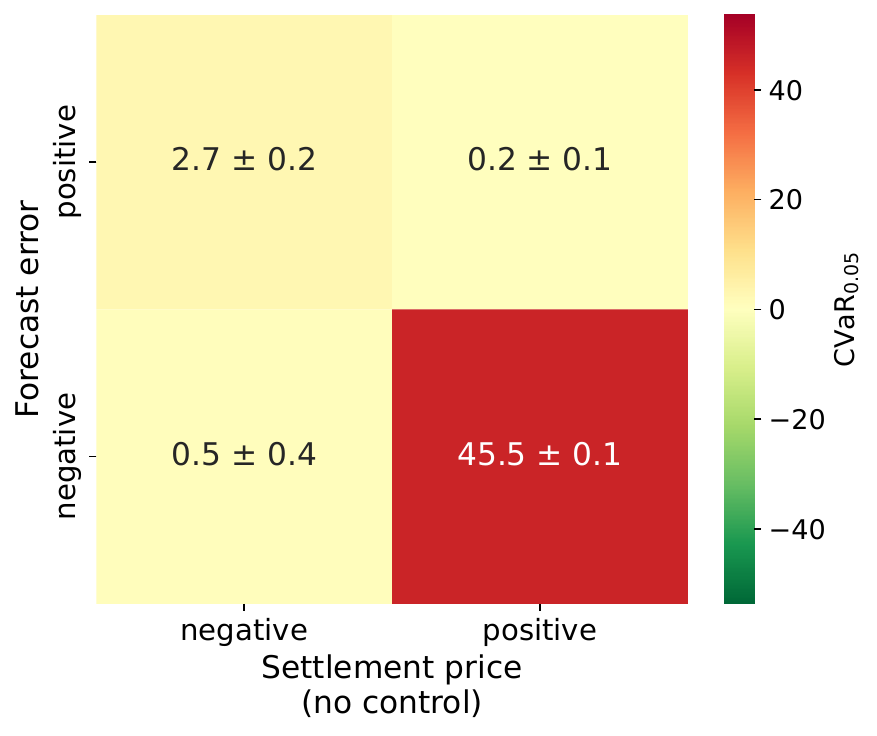}
        \caption{distSAC $\mathrm{CVaR}_{0.01}$}
    \end{subfigure}
    \caption{Out-of-sample conditional value at risk ($\beta=0.05$) for the no-control baseline and the distSAC policies. Positive values represent imbalance costs, but the risk level is defined with respect to the profit distribution to remain consistent with the reward definition.  The results are grouped by the forecast error, i.e., the average difference between the possible power and the trading schedule during the settlement period, and the settlement price without control actions. For the distSAC policies, each statistic is reported as the average $\pm$ standard deviation over five training seeds, the no-control baseline is deterministic and therefore has no standard deviation.}
\end{figure}

\begin{figure}
    \centering
    \includegraphics[width=0.6\linewidth]{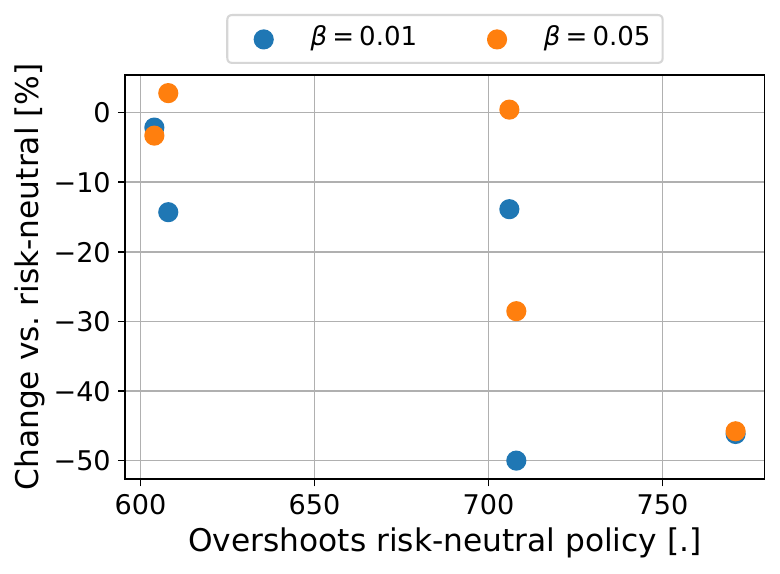}
    \caption{Change in the number of overshoots of the risk-averse policies ($\beta=0.01$ and $\beta=0.05$) relative to the risk-neutral policy, per seed. An overshoot is a settlement period in which curtailment turns a negative price positive. The x-axis shows the number of overshoots of the risk-neutral policy, the y-axis the relative change. Negative values indicate fewer overshoots.}
    \label{fig:placeholder}
\end{figure}

\end{document}